\documentclass[aps,twocolumn,superscriptaddress,amsmath,amssymb]{revtex4}
\usepackage{graphicx}
\usepackage{color}
\usepackage{colordvi}
\usepackage{amsmath}
\begin{document}
\newcommand{\ltwid}{\mathrel{\raise.3ex\hbox{$<$\kern-.75em\lower1ex\hbox{$\sim$}}}}
\newcommand{\gtwid}{\mathrel{\raise.3ex\hbox{$>$\kern-.75em\lower1ex\hbox{$\sim$}}}}

\newcommand{\bea}{\begin{eqnarray}}
\title {Anomaly at $\Delta_{\bf k}$ in the angle-resolved
photoemission spectrum of dirty superconductors}

\author{T.~Dahm}
\email{thomas.dahm@uni-tuebingen.de}
\affiliation{Institut f\"{u}r Theoretische Physik,
Universit\"at T\"ubingen, T\"ubingen, Germany}

\author{P.J.~Hirschfeld}
\email{pjh@phys.ufl.edu}
\author{L.~Zhu}
\email{zly@phys.ufl.edu}
\affiliation{Physics Department, University of Florida, Gainesville,
FL 32611 USA}

\author{D.J.~Scalapino}
\email{djs@vulcan2.physics.ucsb.edu}
\affiliation{Department of Physics, University of California,
Santa Barbara, CA 93106-9530 USA}

\date{\today}

\begin{abstract}{ Elastic forward scattering can lead to an
anomaly in the angle-resolved
photoemission spectrum of the cuprate superconductors. Here
we discuss how this anomaly can be used to provide a measurement
of the superconducting gap $\Delta_{\bf k}$ for {\bf k} values 
away from the Fermi surface. }
\end{abstract}
 \maketitle

In the search to understand the mechanism responsible for pairing
in the cuprate high $T_c$ superconductors, a key signature is
thought to be the momentum dependence of the gap. A gap which has
the simple $(\cos k_x-\cos k_y)$ dependence throughout the
Brillouin zone would imply that the pairing arose from a near
neighbor $Cu$-$Cu$ interaction such as a superexchange coupling.
Additional {\bf k} dependence, involving higher $d_{x^2-y^2}$
harmonics, would imply a more extended spatial
interaction\cite{ref1}, and some angle-resolved photoemission
spectroscopy (ARPES) data on underdoped cuprates
have indeed been used to suggest that the underdoped materials may
have a longer range pairing\cite{Kaminski00}. Information on the
gap function for values of $\bf k$ on the Fermi surface have been
obtained from low temperature transport measurements \cite{LT},
which probe the nodal regions, and from ARPES measurements\cite{Damamodern}.
While the peak at the quasiparticle energy $E({\bf k})$ in the
ARPES energy distribution function (EDC) can in principle be used
to determine $\Delta_{\bf k}$ for $\bf k$ values away from the
Fermi surface, this requires assuming a band structure and
neglecting  self-energy shifts. In addition, the EDC can be broad
and asymmetric because of interactions, making the determination
of the peak position uncertain. Here we discuss an alternative
possibility for determining the {\bf k} dependence of the gap away
from the Fermi surface for general values of {\bf k} based upon
a structure in $A({\bf k},\omega)$ introduced by forward elastic
scattering processes.

One expects that in Bi$_2$Sr$_2$CaCu$_2$O$_8$ (BSCCO) out-of-plane
impurities or disorder will lead to forward elastic
scattering\cite{Varma}. A general discussion of the effect that
such scattering can have on the ARPES spectrum was previously
given\cite{elastic,elastic2}. Here we will focus on one aspect of the
scattering which leads to an anomalous structural feature in the
ARPES spectral EDC at an energy equal to $\Delta_{\bf k}$.  To
illustrate the idea, we begin by considering just the effect of
strong forward elastic scattering where the Nambu self-energies
can be approximated by
\begin{equation}
\Sigma^{''}_0({\bf k},\omega)=-\Gamma_0({\bf k})
\frac{|\omega|}{\sqrt{\omega^2-\Delta_{\bf k}^2}}
\label{sig0}
\end{equation}
\begin{equation}
\Sigma^{''}_1({\bf k},\omega)=-\Gamma_0({\bf k}) \frac{\Delta_{\bf
k}{\rm sgn}\omega }{\sqrt{\omega^2-\Delta_{\bf k}^2}}
\label{sig1}
\end{equation}
\begin{equation}
\Sigma^{''}_3({\bf k},\omega)\cong0.
\end{equation}
The conditions for this approximation to be valid were discussed
in \cite{elastic}. Here $\Gamma_0({\bf k}) $ is the normal state
scattering rate due to out-of-plane impurities. Note, that expressions
(\ref{sig0}) and (\ref{sig1}) possess square-root divergences at the 
gap edge $\omega=\pm \Delta_{\bf k}$. Using these
self-energies, the one-electron spectral weight can be written as
\cite{elastic,Rarkeu}
\begin{equation}
A({\bf k},\omega)=-\frac{1}{\pi}{\rm Im}\{\frac{\omega Z({\bf
k},\omega)+\epsilon_{\bf k}}{(\omega^2-\Delta_{\bf k}^2)Z^2({\bf
k},\omega)-\epsilon_{\bf k}^2}\}
\label{four}
\end{equation}
where $Z({\bf k},\omega)=1+i\Gamma_0({\bf k})\, {\rm
sgn}\,\omega/\sqrt{\omega^2-\Delta_{\bf k}^2}$. In the following
we will set
\begin{equation}
\epsilon_{\bf k}=-2t(\cos k_x+\cos k_y)-4t'\cos k_x\cos k_y-\mu
\end{equation}
with $t'/t=-0.35$ and $\mu/t=-1$, giving the Fermi surface shown
in Fig 1. We will also assume for the purpose of illustration that
\begin{equation}
\Delta_{\bf k}=\frac{\Delta_0}{2}(\cos k_x-\cos k_y)
\end{equation}
with $\Delta_0=0.2t$.

   Fig.~2(a) shows a plot of $A({\bf k},\omega)$ for several values of the
normal state forward scattering rate $\Gamma_0({\bf k})$ for ${\bf
k}=(0.64\pi,0)$ [the start of the A-cut shown in Fig.~1].  As
expected, there is a broadened quasiparticle peak at
$\omega=-\sqrt{\epsilon_{\bf k}^2+\Delta_{\bf k}^2}$.
   However, in addition there is a square-root anomaly as $\omega$
   approaches $-\Delta_{\bf k}$. From eq.~\eqref{four} one finds that for
{\bf k} away
   from the Fermi surface with $|\epsilon_{\bf k}|>\Delta_{\bf k}$,
   \begin{equation}
   A({\bf k},\omega\rightarrow
   -\Delta_{\bf k})\simeq\frac{\Delta_{\bf k}\Gamma_0({\bf k}) }{\pi\epsilon^2_{\bf k}}\frac{1}{\sqrt{\omega^2-\Delta_{\bf k}^2}} \label{eq7}
   \end{equation}
   Thus, while the strength of the square-root anomaly caused by the
   forward scattering decreases as one goes deeper below the Fermi
   surface, the structure remains, and should be observable if the
scattering rate $\Gamma_0({\bf k})$ is sufficiently
   large.
\begin{figure}[tb]
\begin{center}
\leavevmode
\includegraphics[width=0.5\columnwidth]{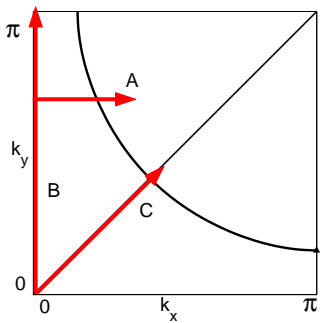}
\caption{Fermi surface in one quadrant of the first Brillouin zone
for $t^\prime/t=0.35$ and $\mu/t=-1$ corresponding to a filling
$n=.83$. Also shown are various momentum cuts A, B, and C for
which EDC spectra are shown in Figures 3 and 4.} \label{fig:fig1}
\end{center}
\end{figure}

   In general, out-of-plane impurities will give rise to a
   scattering potential characterized by a finite range
   $\kappa^{-1}$, and a more complete, self-consistent treatment
   of the self-energy due to impurity scattering is required.
Here, we follow the approach taken in \cite{elastic} in which a
simple exponential scattering potential $V_0e^{-\kappa r}$ was
treated within a self-consistent Born approximation. In this case,
the scattering strength depends upon both {\bf k} and $\omega$.
For $\omega=-\Delta_{\bf k}$, the scattering strength is
suppressed when $|\epsilon_{\bf k}|$ is greater than
$\epsilon_\kappa \equiv v_F\kappa$ due to phase space
restrictions. Nevertheless, as shown in Fig.~2(b), the anomaly
continues to occur at $\omega = -\Delta_{\bf k}$, although the
structure can be
  broadened somewhat.
   Here, we have set $\kappa =1$, measured in inverse units of the Cu-Cu
spacing, and chosen the strength of the out-of-plane impurity scattering
potential and the impurity concentration to fix the normal state scattering
rate $\Gamma_0$ equal to $2\Delta_0$, $\Delta_0$, and $\Delta_0/2$.

\begin{figure}[h]
\begin{center}
\leavevmode
\includegraphics[width=\columnwidth]{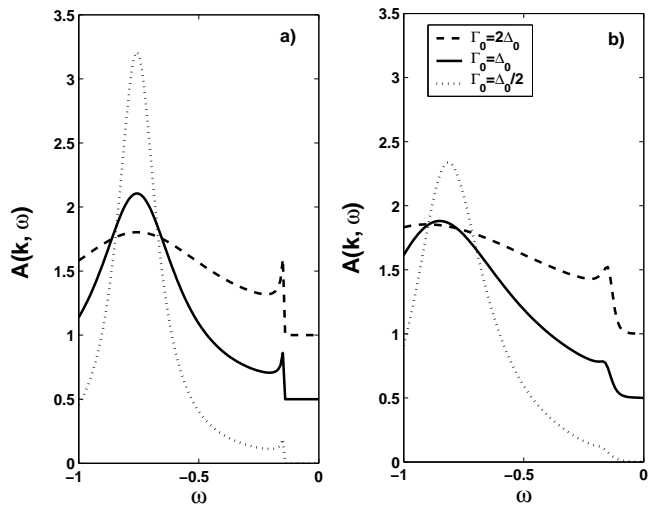}
\caption{ a) Approximate spectral function $A({\bf k},\omega)$
from Eq (4) versus $\omega$ for ${\bf k}=(0.64\pi,0)$ with
$\Gamma_0=2\Delta_0$ (dashed curve), $\Delta_0$ (solid curve) and
$\Delta_0/2$ (dotted curve), with $\Delta_0=0.2t$. b) shows the
numerical result obtained from the self-consistent Born
approximation as described in Ref. \cite{elastic}, with $\kappa=1$
and the same scattering rates. Curves are offset by 0.5 with respect
to each other and $\omega$ is given in units of $t$.}\label{fig:fig2}
\end{center}
\end{figure}

  In general, many-body interactions give rise to a
    gap function which depends upon both ${\bf k}$ and $\omega$.
   However, at low temperatures when $\omega$ is equal to the real
   part of the gap at the gap edge i.e. $
   \Delta_{\bf k}={\rm Re}[\Delta({\bf k},\omega=\Delta_{\bf k})]$, the
broadening due to inelastic scattering vanishes as $T^3$ 
\cite{Quinlan} and the
imaginary part of the gap is determined by elastic impurity
scattering only, i.e. the anomaly at $\Delta_{\bf k}$ is stable against
many-body interactions. Thus, out-of-plane forward impurity scattering
gives rise to the anomaly and in-plane isotropic impurity
scattering leads to only a small broadening for low energies near
the Fermi level. In our units, a typical value for the broadening
due to in-plane scattering is of order .02t, a typical value of
the in-plane scattering estimated from fits to STM studies
\cite{ZAH03}.  It is irrelevant for the anomaly at the gap edge
and other features at higher binding energies.

\begin{figure}[tb]
\begin{center}
\leavevmode
\includegraphics[width=\columnwidth]{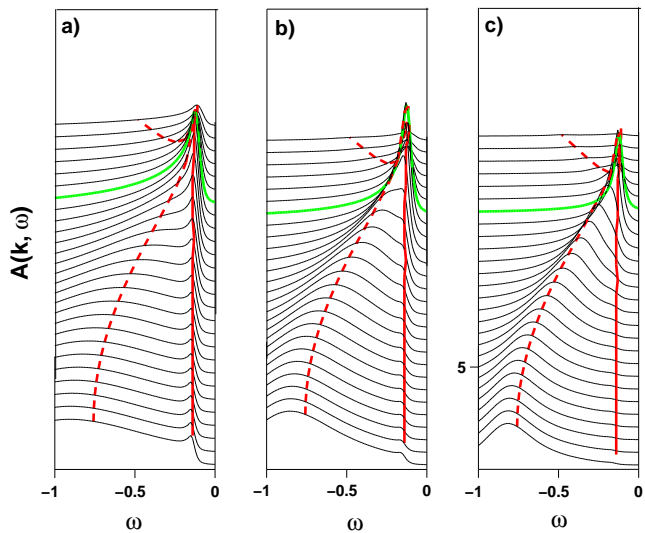}
\caption{a-c $A({\bf k},\omega)$ versus $\omega$ for ${\bf k}$
along the A cut shown in Figure 1 with a momentum separation $\Delta {\bf
k}=0.035$.  These plots illustrate how the
anomaly varies for $\kappa=1$ and  different values of the the
scattering strength: (a)$\Gamma_0=2\Delta_0$; (b)
$\Gamma_0=\Delta_0$; (c) $\Gamma_0=\Delta_0/2$, with
$\Delta_0=0.2t$. The solid red curve indicates the $\Delta_{\bf
k}$ anomaly, while the dashed red curve follows the quasiparticle
dispersion.  Fermi surface crossing spectra are colored green.}
\label{fig:fig3}
\end{center}
\end{figure}

In order to illustrate the type of behavior one is looking for, in
Fig.~3 we have plotted $A({\bf k}, \omega)$ versus $\omega$ for a
set of ${\bf k}$ values taken along the A-cut shown in Fig.~1.
Here, $\kappa=1$ and results including a self-consistent Born
approximation treatment of the forward scattering, together with
an in-plane constant scattering rate of 0.02$t$, are shown for
$\Gamma_0=2\Delta_0$, $\Delta_0$, and $\Delta_0/2$. In Fig.~3, the
solid red curve follows the $\Delta_{\bf k}$ anomaly while the dashed
red curve indicates the quasiparticle energy
$-\sqrt{\epsilon^2_{\bf k} + \Delta^2_{\bf k}}$. The various
$A({\bf k}, \omega)$ spectra are offset for the different {\bf k}
values which are taken at a momentum separation of $\Delta {\bf k}
= 0.035$ along the A-cut shown in Fig.~1.

As seen, the strength of the anomaly \cite{ref2} depends upon the out-of-plane
impurity scattering rate $\Gamma_0$. For $\Gamma_0\gtwid \Delta_0$, there
is a clear anomaly which occurs at $\omega = - \Delta_{\bf k}$. Here, we have
taken $\Delta_{\bf k} = \frac{\Delta_0}{2} (\cos k_x-\cos k_y)$ for
illustration. In general, the anomaly occurs at $\Delta_{\bf k}={\rm Re}[\Delta({\bf
k}, \omega=\Delta_{\bf k})]$ and the locus of this anomaly allows one to
determine the {\bf k}-dependence of the gap for general values of {\bf k}.

\begin{figure}[tb]
\begin{center}
\leavevmode
\includegraphics[width=\columnwidth]{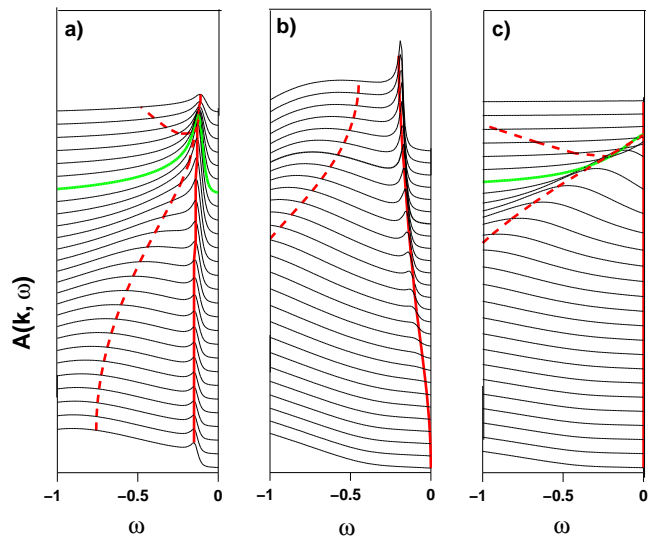}
\caption{a-c $A({\bf k},\omega)$ versus $\omega$ for ${\bf k}$
taken at intervals of $\delta{\bf k}=0.0385,0.12,0.065$ along the
A, B, C cuts shown in Figure 1, for $\Gamma_0 = 2\Delta_0$ and
$\kappa =1$ . The solid red curve shows the locus of the anomaly
which follows $\Delta_{\bf k}=\frac{\Delta_0}{2}(\cos k_x-\cos
k_y)$. The dashed red curve follows the quasi-particle dispersion
$-\sqrt{\epsilon_{\bf k}^2+\Delta_{\bf k}^2}$, and the green curve
represents the EDC at the Fermi surface crossing point.}
\label{fig:fig4}
\end{center}
\end{figure}

Further results for $A({\bf k}, \omega)$ taken along the three momentum
cuts A, B, and C are illustrated in Fig.~4. Here $\kappa =1$ and $\Gamma_0$ has
been set to $2\Delta_0$. As before, the solid red curve gives the locus of
$-\Delta_{\bf k}$, while the dashed red curve follows the broadened
quasiparticle dispersion $-\sqrt{\epsilon^2_{\bf k} + \Delta^2_{\bf k}}$.
As seen along the B cut, the strength of the gap anomaly decreases as
$\Delta_{\bf k}$ decreases. It is also suppressed when $|\epsilon_{\bf k}| >
\epsilon_\kappa$ due to phase space restrictions associated with the 
forward elastic scattering process.

\begin{figure}[h]
\begin{center}
\leavevmode
\includegraphics[width=\columnwidth]{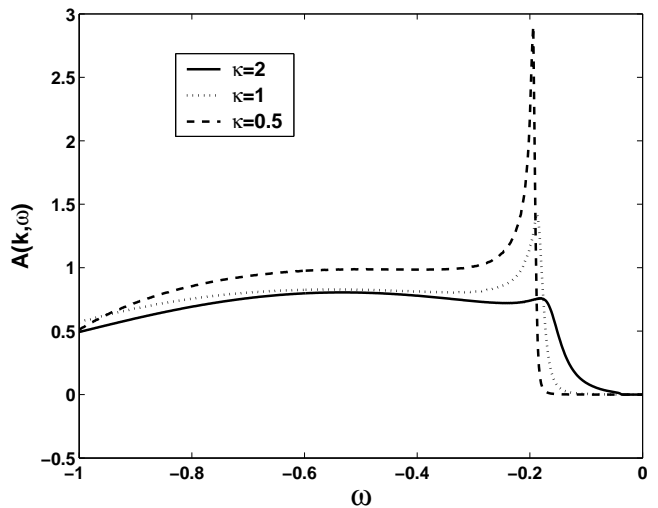}
\caption{  
$A({\bf k},\omega)$ versus $\omega$ for ${\bf k}=(\pi,0)$ with
$\Gamma_0=2\Delta_0$ and three different values of $\kappa=2$ (solid curve), 
$\kappa=1$ (dotted curve) and $\kappa=0.5$ (dashed curve), with $\Delta_0=0.2t$. 
}\label{fig:fig5}
\end{center}
\end{figure}

In Fig.~\ref{fig:fig5} we show $A({\bf k}, \omega)$ at the M point
${\bf k}=(\pi,0)$ for three different values of $\kappa$ in order to
illustrate the influence of the range of the scattering potential
on the anomaly. A rough criterion for the appearance of the anomaly
is that $\kappa$ should become smaller than 1, which means that the
range of the scattering potential should become larger than one
lattice spacing of the CuO$_2$ planes. This is likely to be the case
for out of plane impurities in the cuprates because of their poor
screening.

In the foregoing analysis of the anomaly at $\Delta_{\bf k}$ we
have restricted ourselves to the self-consistent Born approximation.
It would be interesting to study effects beyond Born approximation,
for example by employing the T-matrix approximation as in a recent
work by Rieck et al \cite{Rieck}, to see how this might affect the
strength of the anomaly. The basic conclusion of our work, i.e. the 
existence and position of the anomaly at $\omega=-\Delta_{\bf k}$, 
will not be affected by the overall magnitude of the  impurity
potential, however  (see, e.g. Eq.~(\ref{eq7}) ).  We leave the
full calculation of the effect on ARPES of finite range scatterers
of arbitrary potential for a future study. 

In summary, the enhancement of the {\bf k}-dependent density of
states at the gap edge $\omega = \pm \Delta_{\bf k}$ leads to an
onset anomaly in $A({\bf k}, \omega)$ due to forward elastic
impurity scattering. The locus of this anomaly
provides a direct measure of the {\bf k}-dependent gap
$\Delta_{\bf k}$ at the gap edge.  Strictly speaking, $\Delta
_{\bf k} = {\rm Re}\, \Delta({\bf k}, \omega=\Delta_{\bf k})$ where
$\Delta({\bf k}, \omega)$ is the complex {\bf k} and
$\omega$-dependent gap function.  As noted, at low temperatures
with $\omega=\Delta_{\bf k}$, the imaginary part of the gap
arising from inelastic scattering vanishes as
$(T/\Delta_0)^3$ and in-plane elastic scattering leads to
only a small imaginary contribution.  Since out-of-plane
forward elastic scattering gives rise to this anomaly, one would
like to be able to control the number of  surface defects,
adjusting $\Gamma_0$ to obtain an optimal measurement. One
possibility may be to purposely add impurities to the surface.  It
is also possible that older BSCCO samples may contain sufficient
out of plane disorder to see the effect discussed here with modern
energy and momentum resolution.  Alternatively, as the experiment
proceeds and the surface gradually becomes contaminated, one may
observe the development of this anomaly.

\begin{acknowledgments}

 We would like to thank Z.X.~Shen and his group and S.A.~Kivelson for useful
discussions. Partial support was provided by
NSF grant DMR02-11166 (DJS) and ONR N00014-04-0060 (PJH and LYZ).

\end{acknowledgments}


\begin{thebibliography}{stuffstuffstuff}


\bibitem{ref1}The $\omega$-dependence of the gap is also of interest
since it reflects the retarded nature of the interaction. Here, we
will focus on the question of the {\bf k} dependence of the gap at
a frequency equal to the gap.

\bibitem{Kaminski00} J.~Mesot {\it et al.}, Phys. Rev. Lett. {\bf 83},
840 (1999).

\bibitem {LT} L.~Taillefer, B.~Lussier, R.~Gagnon, K.~Behnia, and H.~Aubin, 
{\sl Phys.~Rev.~Lett.} {\bf 79}, 483 (1997).

\bibitem {Damamodern} A.~Damascelli, Z.~Hussain and Z.X.~Shen, {\sl
Rev.~Mod.~Phys.} {\bf 75}, 473 (2003).

\bibitem {Varma} E.~Abrahams and C.M.~Varma, {\sl Phys.~Nat'l Acad.~Sci.}
{\bf 97} 5714 (2000).

\bibitem{elastic} L.~Zhu, P.J.~Hirschfeld,
and D.J.~Scalapino, {\sl Phys. Rev. B} {\bf 70}, 214503 (2004).

\bibitem{elastic2} D.J.~Scalapino, T.S.~Nunner, and P.J.~Hirschfeld,
 cond-mat/0409204.

\bibitem{Rarkeu} R.S.~Markiewicz, {\sl Phys. Rev. B} {\bf 69}, 214517 (2004).

\bibitem{Quinlan} S.M.~Quinlan, D.J.~Scalapino, and N.~Bulut,
{\sl Phys.~Rev.~B} {\bf 49}, 1470 (1994).

\bibitem{ZAH03} L.~Zhu, W.A.~Atkinson, and P.J.~Hirschfeld,
{\sl Phys.~Rev.~B} {\bf 69}, 060503(R) (2004).

\bibitem{ref2} It also depends upon $\kappa$ and for {\bf k} values deep
inside the Fermi surface the size of the anomaly is suppressed by
$(\epsilon_\kappa/|\epsilon_{\bf k}|)^5$ for the exponential
potential we have used.

\bibitem{Rieck} C.~T.~Rieck, K.~Scharnberg, and S.~Scheffler, cond-mat/0408320.

\end{thebibliography}
\end{document}